# Magnetodielectric behavior in $La_2CoMnO_6$ nanoparticles

**J. Krishna Murthy and A. Venimadhav[a]**

*Cryogenic Engineering Centre, Indian Institute of Technology, Kharagpur-721302, India*

## Abstract

We have investigated magnetic, dielectric and magnetodielectric properties of $La_2CoMnO_6$ nanoparticles prepared by sol-gel method. Magnetization measurements revealed two distinct ferromagnetic transitions at 218 K and 135 K that can be assigned to ordered and disordered magnetic phases of the $La_2CoMnO_6$ nanoparticles. Two dielectric relaxations culminating around the magnetic transitions were observed with a maximum magnetodielectric response reaching 10% and 8% at the respective relaxation peaks measured at 100 kHz under 5T magnetic field. The dc electrical resistivity followed an insulating behavior and showed a negative magnetoresistance; there was no noticeable anomaly in resistivity or magnetoresistance near the magnetic ordering temperatures. Complex impedance analysis revealed a clear intrinsic contribution to the magnetodielectric response; however, extrinsic contribution due to Maxwell-Wagner effect combined with magnetoresistance property dominated the magnetodielectric effect at high temperatures.

*a) Corresponding author: venimadhav@hijli.iitkgp.ernet.in*

# Introduction

Magnetic field induced change in dielectric permittivity namely magnetodielectric (MD) effect or magnetocapacitance originates in both multiferroic and non-multiferroic systems.[1] This functional property can be explored for intelligent sensor, spintronic and functional device applications. In non-multiferroic systems the observed MD effect has been related to various mechanisms like spin-lattice coupling,[2] orbital ordering[3] and strain mediated coupling between electric and magnetic order parameters.[4] However, extrinsic Maxwell-Wagner (MW) effect combined with Magnetoresistance (MR) can also induce significant MD effect in many systems.[5]

Double pervoskite oxides with general formula $A_2B'B''O_6$ (A-rare earth, B' and B''- 3d transition metals) have displayed a wide variety of interesting physical properties with compositional variations.[6, 7] Non-multiferroic, $La_2NiMnO_6$ (LNMO) has gained more attention due to its large MD effect close to room temperature and this was attributed to spin lattice coupling.[6] The isostructural $La_2CoMnO_6$ (LCMO) system has showed a ferromagnetic (FM) transition temperature ($T_c$) ~ 225 K with an insulating behavior.[7] Several reports have shown that the magnetic properties are related to cation ordering in LCMO system which in turn depends on the synthesis conditions.[7-11] It exhibits one or more magnetic transitions; an ordered sublattice with high spin $Co^{2+}$ and $Mn^{4+}$ pair giving a FM transition ~220 K[7, 8] while a disordered sublattice with low spin $Co^{3+}$ and high spin $Mn^{3+}$ invokes a FM transition below 150 K.[9-10] It was reported that in the case of low temperature sintered samples, high $T_C$(~ 220 K) was observed while the samples prepared at high temperature exhibited low $T_C$(~150 k).[11] Recently, Singh *et al,* have reported MD effect in the ordered LCMO thin films near $T_C$ and attributed it to spin-lattice coupling.[8] In the polycrystalline LCMO system, Lin *et al,* have observed 0.8% of MD response at 280 K.[12] Though Raman measurements showed a spin-lattice coupling near the two magnetic transitions in LCMO bulk,[9] there is no systematic study relating the MD effect to the FM ordered/disordered phases of LCMO polycrystalline system. Here, we report the observation of dielectric relaxations around the magnetic ordering temperatures in LCMO nanoparticle system and based on complex impedance analysis, various contributions to the observed MD property have been discussed.

**Experimental**

LCMO nanoparticles sample has been prepared by conventional sol-gel based polymeric precursor route. Reagent grade chemicals such as $La_2O_3$, $Co(NO_3)_2 6H_2O$, $Mn(CH_3COO)_2 4H_2O$ were taken as the precursor materials and they were weighted according to the stoichiometric ratios. Lanthanum oxide ($La_2O_3$) is converted to Lanthanum nitrate solution by dissolving $La_2O_3$ in the nitric acid; later $Co (NO_3)_2 6H_2O$ and $Mn(CH_3COO)_2 4H_2O$ were dissolved in deionised water and they were mixed with Lanthanum nitrate solution. To this precursor mixture ethylene glycol was added while continuously stirring the solution. The solvent was evaporated directly in the temperatures range between 75°C to 150°C using a hot plate until a solid resin is formed. The resin was calcinated at temperatures 600°C for 2 hours in the furnace. For dielectric measurements the powders were pressed into pellets of 10 mm dia and sintered at 600°C for 5 hours. The phase analysis was carried out using Phillips powder x-ray diffractometer (XRD) with Cu-$K_\alpha$ radiation and JEOL JEM2100 high-resolution transmission electron microscope (HRTEM) has been used to estimate the particle size. The temperature dependence of ac susceptibility measurement was done with EverCool Quantum Design SQUID-VSM magnetometer. Dielectric properties were investigated through an Agilent 4294A impedance analyzer with an ac excitation of 500 mV. Conducting silver paste contact was applied on both sides of the sample to make parallel plate capacitor geometry. The temperature dependence of dc electrical resistivity was measured with conventional four probe method. The temperature and the magnetic field variation of dielectric properties were performed using a closed cycle cryogen free superconducting magnet system from 300 K to 8 K.

## Results and Discussion

Fig 1(a) shows the XRD pattern of the LCMO nanoparticles; it shows a single phase with no impurities. The observed reflections were assigned to pseudo tetragonal crystal structure as reported by Dass *et al.*[7] and the crystallite size was estimated to be ~ 32 nm from Debye- Scherrer formula. Fig.1 (b) shows a TEM micrograph of homogeneously distributed LCMO nanoparticles with an average size of ~ 28 nm that is consistence with crystallite size (~ 32 nm) estimated from the XRD pattern. Fig.1(c) shows the HRTEM lattice image with a lattice spacing d ~ 0.29 nm corresponding to (112) plane. The ring pattern in the selective area electron diffraction (SAED) image of the fig.1 (d) represents the polycrystalline nature of sample.

Figure 2 shows the dc electrical resistivity of the LCMO sample measured up to 135 K under the application of 0T and 5T magnetic field. The plot shows an insulating behavior as reported earlier on the bulk LCMO samples.[13] Isothermal field variation of MR % (= $[\frac{\rho_{H(T)} - \rho_{0(T)}}{\rho_{0(T)}}] * 100$ ) at 150 and 220 K is shown in the upper inset to fig.2. We have observed a negative MR and the magnitude of MR% increases linearly with the decrease of temperature under 5T field. The variation in dc resistivity with and without applied magnetic field can be fitted to variable range hopping (VRH) mechanism[14] and its activation energy (E) can be calculated as follows,

$$\rho(T) = \rho_o \exp\left(\frac{B}{T}\right)^{1/4} \quad \text{---------- (1)}$$

Where B $=4E/(K_B T^{3/4})$ and E is the activation energy. The observed zero magnetic field activation energy (E) ~ 150 meV and 180 meV at 150 and 220 K respectively.

Figure 3(a) shows the temperature dependent imaginary part of ac susceptibility ($\chi''$) measured with ac excitation field of 1 Oe at 523 Hz. In the fig. 3(a), the cusp at $T_{C1}$~ 218 K can be assigned to a strong PM- FM phase transition. We have also observed another weak FM transition at $T_{C2}$ ~135 K due to disorder phase of LCMO nanoparticles. These two magnetic transitions at 135 and 218 K are in agreement with the previous reports on bulk and thin film samples of LCMO.[7-11] Here, the first transition at 218 K has been attributed to the FM superexchange interactions of the ordered $Co^{2+}$- $O^{2-}$- $Mn^{4+}$ pair [7-9] and the second transition at 135 K can be assigned to FM vibronic superexchange interaction of intermediate spin $Co^{3+}$ ($t_{2g}^3 e_g^1$)–high spin $Mn^{3+}$ pair.[9-10] The second transition results due to the occurrence of oxygen vacancies during the synthesis of LCMO nanoparticles. The low temperature transitions at ~ 33 K showed the frequency dependence of ac susceptibility indicating a magnetic glassy nature, further details are being investigated. The magnetic field dependence of magnetization at 5K (not shown) has given a saturation magnetization of $M_S$ ~3.68 $\mu_B$/f.u; this is much smaller than the theoretically calculated spin only value of 6.0 $\mu_B$ /f.u. Further, an applied magnetic field of 6T has been found to be insufficient to obtain complete saturation at 5 K. These observations suggest the presence of oxygen vacancies and antisite defects in the LCMO nanoparticle system.

Fig. 3(b) and (c) shows, temperature dependent real part of dielectric permittivity ($\varepsilon'$) and loss tangent (Tan δ) for different frequencies (500-100 kHz). The dielectric permittivity decrease sharply with temperature and exhibited two step like features. Further decrease in temperature shows a saturation value of $\varepsilon' \sim 10$ irrespective of the frequency due to dipolar freezing.[14] The observed two drops in the dielectric permittivity as shown in the fig. 3(b) are accompanied by two relaxation peaks (see fig.3(c)) in high temperature (first relaxation in 150-250 K) and low temperature (second relaxation in 90-160 K) regions respectively. The variation of MD effect [MD (%) = $(\frac{\varepsilon_{5T} - \varepsilon_{0T}}{\varepsilon_{0T}}) * 100$] as a function of temperature for different frequencies at 5T field is shown the fig.3 (d). As shown in the figure, a small value (< 1%) of MD effect near to room temperature (i.e., in paramagnetic region) can be observed. This effect increases with reducing of temperature and shows first maxima around ~ 220 K corresponding to the first dielectric relaxation; further decrease in temperature shows second maxima around 135 K corresponding to the second dielectric relaxation. The MD magnitude of first maxima (20% at 5 kHz and 10% at 100 kHz) is higher when compared to the MD magnitude (10% at 5 kHz and 8% at 100 kHz) of the second maxima. It can be noted from the fig. 3(d) that the difference in the magnitude of two MD peaks is large at low frequencies and this difference reduces becomes almost the same at high frequencies. This implies that the relaxation mechanism seems to play an important role in determining the magnitude of MD effect. Further, both the relaxations have shown frequency dispersion and shifting of relaxation peaks to high temperature implying that the relaxations are thermally activated. The observed relaxation phenomena in LCMO nanoparticles can be fitted to the Arrhenius law given by the equation,

$$\tau = \tau_o \exp\left(\frac{E_a}{k_B T}\right) \quad \text{----------} \quad (1)$$

Where $\tau_o$ is the pre-exponential factor, $E_a$ is the activation energy, and $k_B$ is Boltzmann constant. The experimental dielectric behavior in the high temperature and low temperature dielectric relaxation regions can be fitted to eq. (1) and found a decrease in activation energy from 171 to 158 meV and 92 to 85 meV corresponding to the two relaxations respectively under 5T magnetic field. Further, lowering of relaxation time has also been observed in both the cases under magnetic field. But this high temperature dielectric relaxation value is close to the activation energy measured by VRH model from the dc resistivity. The activation energy (~ 92 meV) corresponding to the low temperature dielectric relaxation matches well with the polaron activation energy (range of 60-90 meV) in wide range of pervoskites.[15,16] From fig.3(a)-(d), it is clear that the observed MD effect occurs in the dielectric relaxation region and the relaxation phenomenon appears to coincide with the two magnetic transitions. Before relating the magnetic ordering with dielectric relaxation one must also understand that a large MD and strong dispersion of MD with frequency can also originate from the extrinsic contributions.[5]

Dielectric permittivity with frequency can shed light on various contributions to the relaxation process. Fig.4 shows the frequency dependence (40-1MHz) of $\varepsilon^{'}$ for selected temperatures under 0T and 5T magnetic fields. The upturn feature in dielectric permittivity observed at high temperatures and low frequency (< 200 Hz) range can be attributed to the formation of schottky type or MIS diode at the sample-contact interface. Two plateau regions in permittivity plot between 100- 260 K in medium and high frequency regions suggest that the present LCMO is an electrically heterogeneous system. At a given temperature, the extrinsic effects of GB/MW effects contribute to the dielectric relaxation in the medium frequency range and intrinsic effects contribute to dielectric relaxation in the high frequency range[14] (as shown in the fig. 4, high frequency relaxation is not very clear due to the experimental limitation beyond 1MHz). For the same temperatures, the frequency dependent dielectric permittivity under the application of 5T magnetic field has showed a similar behavior. But the applied magnetic field shifts the dielectric permittivity drop to high frequency side i.e., relaxation process becomes faster in the presence of magnetic field. This is also evident from the temperature variation of relaxation dynamics (ln$\tau$ vs. 1/T) as shown in inset to fig.3(c).

Complex impedance spectroscopy (CIS) is a good tool to separate the bulk (B) and grain boundary (GB) contributions to the dielectric transport in electrically inhomogeneous materials.[14, 17] Fig.5 shows the complex impedance plane that represents the variation of real (Z') and imaginary (Z") parts of impedance (i.e., Nyquist plots) for the selected temperatures. We have adopted the most general electrical equivalent circuit with the parallel combination of two R-C networks connected in series[17] as shown in inset to fig.5. From this figure it can be noted that for temperatures below 110 K only one semicircle arc is observed and that corresponds to the intrinsic (bulk) contribution. With the increase of temperature appearance of a second semicircular arc can be observed and that can be assigned to the extrinsic (such as GB) effects. As shown in the figure, the measured data fits well with applied circuit model and fitting parameters like $R_B$, $R_{GB}$ and $C_B$, $C_{GB}$ corresponds to resistance and capacitance of bulk and GB can be obtained. The reliability of the above parameters were verified by plotting the M"(imaginary part of electric modulus) and Z"(imaginary part of impedance) versus frequency at different temperatures;[17] corresponding maximum in M" and Z" plots found to match with $f_{max}=1/2\pi R_B C_B$ (Hz) and $f_{max}=1/2\pi R_{GB} C_{GB}$ (Hz) respectively.

The above computed GB and bulk resistances and capacitances have been plotted as a function of temperature for 0T and 5T magnetic fields as shown in the fig. 6(a) and (b); following observations can be made from these two figures: (i) both $R_{GB}$ and $R_B$ increases with decreasing temperature and magnitude of GB resistance is higher than the bulk resistance for all temperatures and there is no anomaly in both $R_{GB}$ and $R_B$ near the two magnetic ordering temperatures; this is consistent with the dc resistivity data as shown in the fig.2, (ii) both GB and bulk capacitance varies similarly with temperature and shows anomalies near the two dielectric relaxations and (iii) in the high temperature region, GB contribution to the MD effect is more than that of the bulk; however, the bulk contribution to MD is present throughout the measured region. This suggests that both intrinsic and extrinsic effects contribute to the observed MD effect and there is strong correlation of dielectric relaxation with the magnetic ordering. In double perovskites,[18] intrinsic spin lattice coupling has been found contribute to the MD behavior; in fact, Troung *et al*,[9] reported that temperature dependent of Raman study in LCMO disordered microcrystals, which confirms that softening of phonon modes at both the FM transitions. Observation of similar MD peak values at the two relaxations for high frequencies (fig 3(d)) and intrinsic contribution to the MD from the impedance analysis (in fig.6(b)) clearly suggests there is spin lattice contribution to the MD below the ferromagnetic ordering temperatures. Supporting to this, we have also observed $\varepsilon_{MD} = \gamma\, M^2$ in the PM-FM transition region [2] (~ 220 K, figure not shown). But large extrinsic effect to MD is also obvious from the impedance analysis. From the fig 6 (a), it is clear that its contribution is high at the high temperature relaxation region and decreases with temperature. Formation of large amount of defects due to oxygen vacancies during the nanoparticles synthesis can destroy the long range cationic ordering [19] and this can lead to canted arrangement of spins. Magnetic data also suggests that the antisites and oxygen vacancies are responsible for the low saturation magnetization. Such a spin disorder within the grains and GB regions can strongly responds to the magnetic field with an enhanced conduction. The close match of activation energies from the VRH model and Arrhenius model near high temperature dielectric relaxation explains the magnetic field driven disorder conductivity; hence the observed MD effect in this region is dominated by the MW model combined with MR mechanism. This still does not explain why the extrinsic $C_{GB}$ shows anomaly at the ordering temperatures; perhaps the dielectric response of the percolation of FM domains (magnetically ordered regions in the background matrix) behaves as MW media at the magnetic transition temperatures. At mid frequencies this effect is dominant and hence the MD peak values are very different (fig 3(d)). Further study on B site ordered LCMO bulk system is necessary to know the individual contributions of the intrinsic and extrinsic effects to the MD property.

In summary, we have prepared polycrystalline LCMO nanoparticles with size ~ 28 nm and studied its magnetic, dielectric, MD and MR properties. A large MD response of 10% and 8% at 100 kHz was observed at the dipolar relaxation peaks. Results demonstrate that both resistance and dielectric permittivity respond to the applied magnetic field, but only dielectric permittivity shows strong relaxation and MD response at the magnetic ordering temperatures. Impedance analysis supports that both intrinsic and extrinsic effects account for the observed MD response. At high temperature relaxation region MD response majorly originated from the extrinsic effects like MW capacitor model combined with MR.

# Acknowledgements

This work was supported by DST (SR/FTP/PS-36/2007) and the authors also acknowledge IIT Kharagpur for funding VSM SQUID magnetometer and DST, New Delhi for FIST grant for establishing cryogen free high magnetic field facility. Krishnamurthy thanks CSIR-UGC, Delhi for JRF.

# Figure captions:

Fig. 1: (a) XRD pattern (b) TEM image; inset shows histogram of particles size (C) HRTEM lattice image and (d) SAED pattern of LCMO nanoparticles.

Fig. 2: Temperature variation of electrical resistivity ($\rho$) with 0T and 5T fields; inset (a) shows field dependence of MR (%) at 150K and 200K and (b) shows the fitting to VRH model.

Fig. 3: (a) Temperature variation of imaginary component ($\chi''$) of ac magnetic susceptibility; (b) and (c) shows Temperature dependent of dielectric permittivity ($\varepsilon'$) and loss tangent (Tan$\delta$) respectively for different frequencies; inset to fig.2(c) shows ln$\tau$ vs. 1/T for 0T and 5T for both low temperature and high temperature dielectric relaxation regions; solid line indicates fitting to eq. (2) (d) Temperature dependent of MD effect under 5T for different frequencies.

Fig. 4: Frequency variation (40-1MHz) of $\varepsilon'$ for different temperatures under 0T and 5T; solid and open symbols indicates 0T and 5T field respectively.

Fig. 5: Z' versus Z" plots at different temperatures. Inset shows the equivalent circuit model.

Fig.6: (a) shows temperature dependent GB capacitance ($C_{GB}$) its GB resistance (in the inset) in 0T and 5T (b) shows temperature dependent bulk ($C_B$) capacitance and its resistance (in the inset) in 0T and 5T.

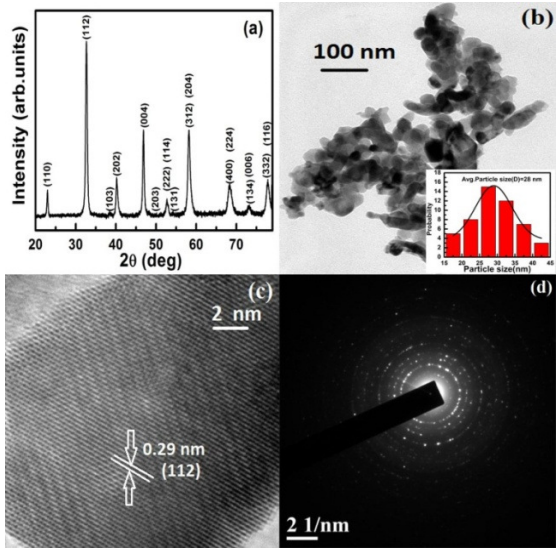

Fig. 1

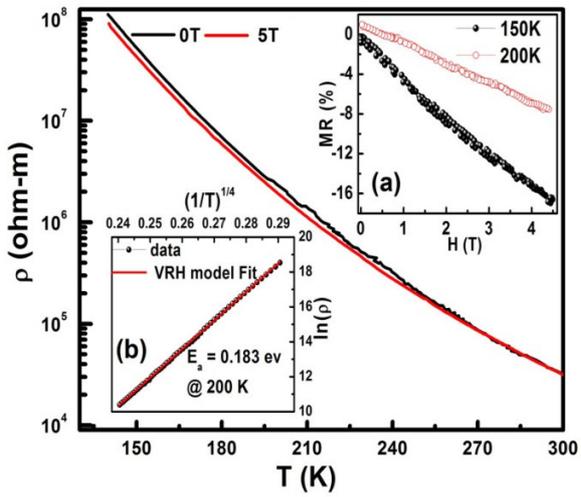

Fig. 2

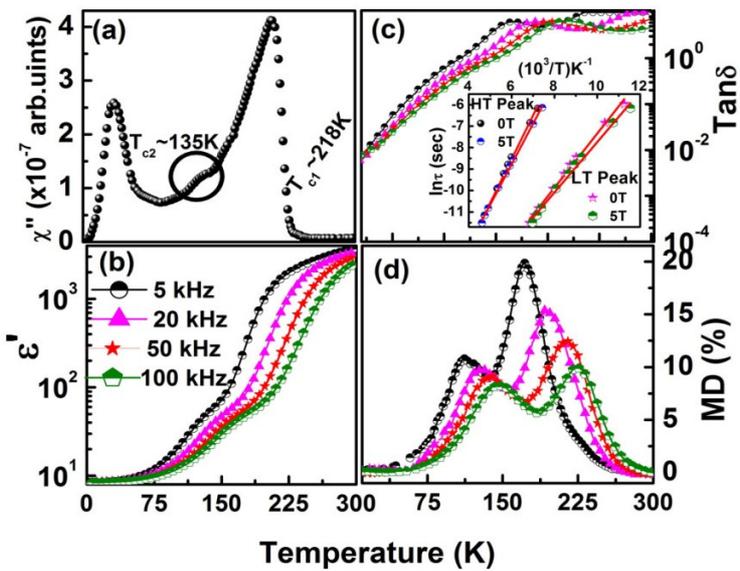

Fig. 3

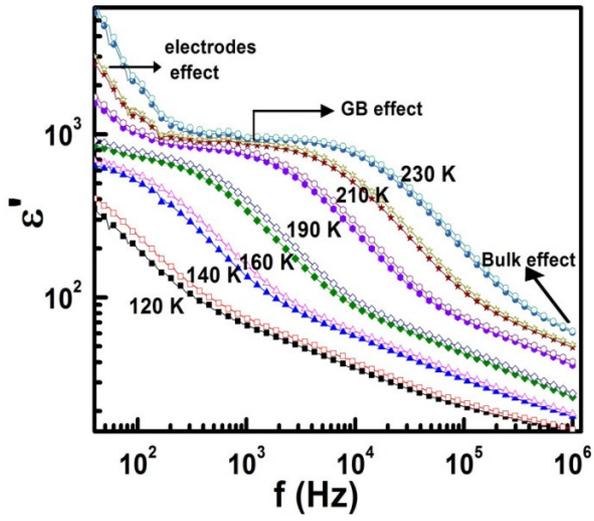

Fig. 4

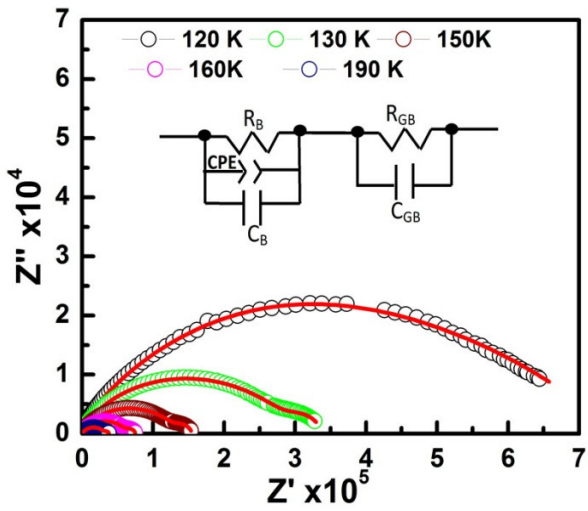

Fig. 5

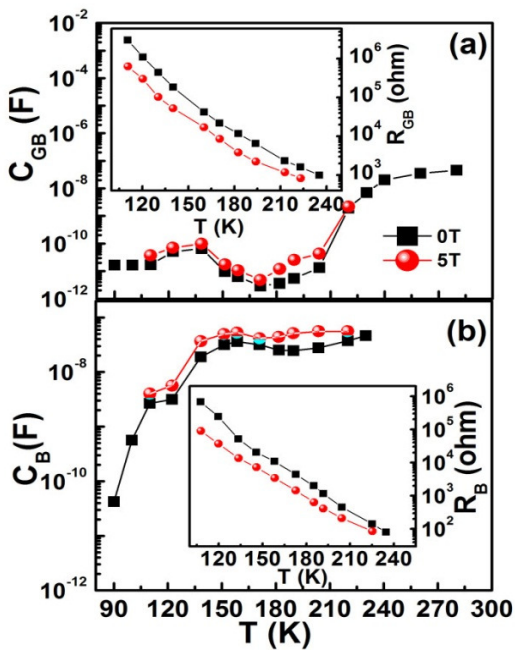

Fig. 6